\documentclass[aps,prd,twocolumn,superscriptaddress,10pt,nofootinbib]{revtex4-2}

\usepackage{tabularx}
\usepackage{graphicx}
\usepackage{amsmath}
\usepackage{amssymb}
\usepackage{comment}
\usepackage[dvipsnames]{xcolor}
\usepackage{xspace}
\usepackage{multirow}
\usepackage{natbib}
\usepackage{hyperref}
\usepackage{url}
\usepackage{booktabs}
\usepackage{multirow}
\usepackage[normalem]{ulem}
\usepackage{orcidlink}

\newcommand\mnras{Monthly Notices of the Royal Astronomical Society}

\renewcommand{\selectlanguage}[1]{}

\begin{document}

\title{Detection and parameter estimation of supermassive black hole ringdown signals using a pulsar timing array}

\author{Xuan Tao\orcidlink{0009-0000-6875-7411}}
\affiliation{National Gravitation Laboratory, MOE Key Laboratory of Fundamental Physical Quantities Measurements, Department of Astronomy and School of Physics, Huazhong University of Science and Technology, Wuhan 430074, China}

\author{Yan Wang\orcidlink{0000-0001-8990-5700}}
\email{ywang12@hust.edu.cn}
\affiliation{National Gravitation Laboratory, MOE Key Laboratory of Fundamental Physical Quantities Measurements, Department of Astronomy and School of Physics, Huazhong University of Science and Technology, Wuhan 430074, China}

\author{Soumya D.~Mohanty\orcidlink{0000-0002-4651-6438}}
\email{soumya.mohanty@utrgv.edu}
\affiliation{Department of Physics and Astronomy, University of Texas Rio Grande Valley, Brownsville, Texas 78520, USA}
\affiliation{Department of Physics, IIT Hyderabad, Kandai, Telangana-502284, India}

\begin{abstract}
Gravitational wave (GW) searches using pulsar timing arrays (PTAs) are commonly assumed to be limited to a GW frequency of $\lesssim 4\times 10^{-7}$Hz given by the Nyquist rate associated with the average observational cadence of $2$ weeks for a single pulsar. However, by taking advantage of asynchronous observations of multiple pulsars, a PTA can detect GW signals at higher frequencies. This allows a sufficiently large PTA to detect and characterize the ringdown signals emitted following the merger of supermassive binary black holes (SMBBHs), leading to stringent tests of the no-hair theorem in the mass range of such systems. Such large-scale PTAs are imminent with the advent of the FAST telescope and the upcoming era of the Square Kilometer Array (SKA). To scope out the data analysis challenges involved in such a search,  we propose a likelihood-based method coupled with Particle Swarm Optimization and apply it to a simulated large-scale PTA comprised of $100$ pulsars, each having a timing residual noise standard deviation of $100$~nsec, with randomized observation times. Focusing on the dominant $(2,2)$ mode of the ringdown signal, we show that it is possible to achieve a $99\%$ detection probability with a false alarm probability below $0.2\%$ for an optimal signal-to-noise ratio (SNR) $>10$. This corresponds, for example, to an equal-mass non-spinning SMBBH with an observer frame chirp mass $M_c = 9.52\times10^{9}M_{\odot}$ at a luminosity distance of $D_L = 420$ Mpc. 
\end{abstract}

\maketitle

\section{Introduction}\label{Sec:intro}
Gravitational Wave (GW) detection in the $\approx [10^{-9}, 10^{-7}]$~Hz band is being pursued by several Pulsar Timing Array (PTA) consortia with observations spanning more than a decade. These include the Parkes PTA (PPTA)~\cite{manchester2013parkes,kerr_parkes_2020}, the European PTA (EPTA)~\cite{kramer_european_2013,chen2021common}, the North American Nanohertz Observatory for Gravitational Waves (NANOGrav)~\cite{McLaughlin_2013,alam_nanograv_2020}, and the Indian PTA~(InPTA)~\cite{joshi_precision_2018,tarafdar_indian_2022}. 
These PTAs further collaborate and exchange data under the umbrella of the International PTA (IPTA)~\cite{verbiest2016international}, with the latest IPTA data release DR2~\cite{antoniadis_international_2022} containing observations of 65 millisecond pulsars (MSPs). In addition, the Chinese PTA~(CPTA)~\cite{lee2016prospects}, utilizing the Five-hundred-meter Aperture Spherical Telescope (FAST) has started timing observations of $57$ MSPs over the last 5 years, among which the timing accuracy of about 35 millisecond pulsars has reached about 100 ns. 
The MeerKAT array in South Africa~\cite{jonas_meerkat_2018}, a precursor to the Square Kilometer Array (SKA)~\cite{carilli_motivation_2004}, has recently released the second data set for the MeerKAT PTA (MPTA)~\cite{miles_meerkat_2025}, covering 4.5 years of observations and containing 83 MSPs with a band-averaged median timing precision of approximately 0.5 $\mu$s.

All the major consortia in the IPTA, as well the CPTA, have recently announced the discovery of a potential stochastic GW background (GWB) signal~\citep{antoniadis_second_2023,reardon_search_2023,agazie_nanograv_2023,xu_searching_2023}. 
Pairwise correlation of the timing residuals from the pulsars in these PTAs reveals a signal that is coherent across the arrays but still lacks strong statistical evidence, with detection significance varying between $\approx 2\sigma$ and $4.6\sigma$, for the quadrupolar characteristics consistent with the Hellings-Downs curve~\cite{hellings_upper_1983} expected from an isotropic GWB. This evidence will likely get stronger as more data is accrued in the future. The recovered amplitude of the common uncorrelated red noise based on the second data release of the MPTA is larger than those reported in the PTAs above~\cite{miles_meerkat_2025}.

It is most likely that the observed GWB signal arises from the population of supermassive binary black holes (SMBBHs) that are expected to form in the merger of galaxies. However, other cosmological origins of the GWB, such as inflation, cosmic strings, and phase transitions, cannot yet be excluded by the present data~\cite{agazie_nanograv_2023}. If the GWB arises predominantly from the SMBBH population, it is natural to expect that future observations will reveal the presence of individual SMBBH sources that stand out from the GWB.
The ever increasing duration of observations by the major existing PTAs as well as the massive expected increase in PTA size and timing precision with FAST and the upcoming SKA~\citep{smits_pulsar_2009,janssen_gravitational_2014} will significantly enhance our observational capabilities in this respect~\cite{2017PhRvL.118o1104W}. 

It is a commonly stated assumption in conventional PTA analyses that the highest detectable GW frequency is limited by the Nyquist rate associated with the observational cadence for a single pulsar~\citep{eptastochastic2015,2015Sci...349.1522S,2016ApJ...821...13A,2018ApJ...859...47A,2014MNRAS.444.3709Z,2016MNRAS.455.1665B, 2019ApJ...880..116A}. Using a typical average observational cadence of two weeks, this limiting frequency is $f_{\rm sp} \approx 4 \times 10^{-7}$~Hz. 
Several efforts have been made~\citep{yardley_sensitivity_2010,yi_limits_2014,perera_improving_2018,dolch_single-source_2016} to extend the high-frequency limit for resolvable GW sources to frequencies greater than~$1\mu$Hz, but they all rely on high-cadence observations of a small number of individual pulsars and cannot be scaled up to a large PTA.
However, it has been shown~\cite{wang_extending_2021} that with the present asynchronous observations of multiple pulsars, the high frequency reach of a PTA for resolvable sources is not limited by $f_{\rm sp} $ but by $ f_{\rm PTA}\leq N_p\times f_{\rm sp} $, where $ N_p $ is the number of pulsars in the array. Thus, using a staggered scheme of interleaved pulsar timing observations, future PTAs with $ N_p \sim O(10^3) $ should be capable of detecting strongly evolving sources at significantly higher frequencies. 
In fact, it may even be possible to detect the dominant $l = 2, m = 2$ mode ringdown signal from the merged black hole at the end of an SMBBH inspiral: for example~\cite{wang_extending_2021}, such a signal could be seen with an optimal signal-to-noise ratio (SNR) of $10$ from a system with a pre-merger rest frame chirp mass of $ \mathcal{M} \lesssim 2 \times 10^{10} M_{\odot} $ at a luminosity distance of  $D_L \lesssim 1.32$~Gpc with an SKA-era PTA containing $N_p = 1000$ pulsars with timing residual noise standard deviation of $100$ nsec per pulsar. 

As with the detection of ringdown signals from smaller mass systems by ground- and space-based detectors, the detection of such signals from merged SMBBHs will be of immense importance across astrophysics and cosmology. It will allow the measurement of the mass and spin of the resulting supermassive black hole, filling the gaps left by other modes of GW observation~\cite{kamaretsos_black-hole_2012,PhysRevD.90.124032}. Additionally, information complementary to that obtained from the inspiral and merger phases will be derived regarding the total mass and mass ratio of the progenitor black holes~\cite{PhysRevD.102.084052} and the orbital configuration of the binary~\cite{PhysRevD.100.084032}.
This information is crucial for the understanding of the formation channel and history of supermassive black holes and the final products of their mergers. Furthermore, ringdown observations would allow tests of general relativity in the strong-field regime for extremely high mass systems~\cite{yunes_gravitational-wave_2013}, e.g., testing the no-hair theorem to an accuracy level of a few percent~\cite{wang_extending_2021}, opening up avenues for discovering new physics~\cite{askar_black_2019}. 

The prospects outlined above for ringdown signal detection using a PTA with staggered sampling have so far been entirely based on considerations of $f_{\rm PTA}$ and the frequency of the signal~\cite{wang_extending_2021}, not on an actual data analysis method. In this paper, we correct this shortcoming by proposing a new data analysis method and characterizing its performance on simulated large-scale PTA data that includes the feature of asynchronous and irregularly spaced observation times. This allows us to quantitatively characterize, for the first time, the prospects of detecting and estimating SMBBH ringdown signals with future large-scale PTAs. The proposed method builds upon the earlier work on the detection and estimation of resolvable continuous wave signals in~\cite{2014ApJ...795...96W, 2015ApJ...815..125W} that combines the Generalized Likelihood Ratio Test (GLRT) with Particle Swarm Optimization (PSO)~\cite{488968}. Following this approach, the extrinsic parameters of the ringdown signal are estimated analytically as in the $\mathcal{F}$-statistic~\cite{jaranowski_data_1998}, while the intrinsic parameters are estimated numerically using PSO.

We demonstrate the method on ringdown signals that contain only the dominant $(2,2)$ mode and a simulated PTA comprised of $100$ pulsars with the standard deviation of the pulsar timing noise being $100$~nsec for each. We consider several different sets of intrinsic parameters, such as the sky location and the frequency of the ringdown source, and study the performance of the method as a function of SNR. 
As an example, we find that for an SNR $= 10$ equal-mass non-spinning SMBBH with a chirp mass $M_c = 9.52\times10^{9}M_{\odot}$ corresponding to frequency $f = 2f_{\rm sp}$ located at a distance of $420$~Mpc, our method achieves a detection probability of $99\%$ at a false alarm probability of less than $0.2\%$. In addition, it yields localization errors with standard deviations of approximately $\sigma_\alpha \approx 5^\circ$ and $\sigma_\delta \approx 4^\circ$, where $\alpha$ and $\delta$ are the right ascension and declination of the SMBBH. 

The rest of the paper is organized as follows. In Sec.~\ref{sec:Ringdown waveform}, we briefly describe the ringdown waveform and the PTA response used in our simulations. Sec.~\ref{sec:Detect} discusses the GLRT statistic used for the detection and estimation of a ringdown signal. It also provides an overview of PSO and the parameter settings for it. 
The simulation setup and results are described in Sec.~\ref{sec:SimResults}.
Discussions and conclusions are in Sec.~\ref{Sec:conc}.


\section{Ringdown waveform}\label{sec:Ringdown waveform}
In this section, we present expressions for the timing residuals induced by GWs from an SMBBH ringdown.
The two GW polarizations $h_{+,\times}$ emitted by a perturbed Kerr black hole can be decomposed into a superposition of damped oscillations called quasi-normal modes. 
The angular distribution of each mode is given by the spin-weighted spherical harmonic ${}_{-2} Y_{l,m}(\theta, \phi)$, with $l = 2, 3, \ldots$ and $-l \leq m \leq l$, while the frequency and damping time of the oscillation are uniquely determined by mass and angular momentum of the resultant black hole after the merger of two black holes. 
Here, $(\theta,\phi)$ denote the polar and azimuthal angles in the source frame of the BH with $\theta \in [0, \pi]$, $\phi \in [0, 2\pi)$ and $\theta=0$ corresponding to the direction of the BH spin. 
Since overtones other than the fundamental mode ($n = 0$) are not excited with significant amplitudes and have shorter damping times, we will consider only the fundamental mode in the current work, although this may lead to biased estimates of the black hole mass and spin in high SNR limit~\cite{PhysRevD.97.044048}. 

Thus, we have 
\begin{align}\label{eq:ringdown}
h(t; \theta, \phi, M, j) &= h_{+} - {\rm i}h_{\times} \nonumber \\
&= \sum_{l=2}^{\infty}\sum_{m=-l}^{l} \! {}_{-2} Y_{l,m}(\theta, \phi) h_{l,m}(M,j,t)\;,
\end{align}
where $M$ is the final black hole mass, and $j$ is the dimensionless spin parameter of the final black hole, respectively. 
In the following, we focus on the $l=m=2$ mode, which is believed to be the most dominant after the black hole perturbation~\cite{flanagan_measuring_1998}.
Starting from the definition of spin-weighted spherical harmonics \cite{goldberg1967spin}, for the $l=m=2$ mode, we get
\begin{equation}\label{eq:Y22}
{}_{-2}Y_{2,2}(\theta, \phi) = \sqrt{\frac{5}{64 \pi}}\left(1+\cos\theta\right)^2 e^{{\rm i} 2 \phi}\;. 
\end{equation}
The corresponding ringdown waveform can be expressed as
\begin{eqnarray}
h_{(2,2)}(t; \theta, \phi, M, j)& = & h_{(2,2),+} - {\rm i} h_{(2,2),\times}  \nonumber \\
&=&\zeta {}_{-2}Y_{2,2}(\theta, \phi) \times \nonumber \\
&& e^{- (t-t_{0})/\tau -{\rm i}2\pi f_{2,2}t  + {\rm i}\varphi_{0} } \;. 
\label{eq:waveform}
\end{eqnarray}
Here, $\zeta  \equiv \mathcal{A}_{2,2}M/{D_L}$ denotes the overall amplitude, $\mathcal{A}_{2,2}$ is the amplitude, $D_L$ is the luminosity distance from the source to the observer,  $t_0$ is the starting time of the ringdown waveforms, $\tau$ is the damping time, $f_{2,2} = \omega_{2,2}/2\pi$ is the frequency of the fundamental overtone in the $(2,2)$ mode, and $ \varphi_{0} $ is the initial phase. 
Similarly, the waveforms $ h_{(2,-2),+,\times} $ can be derived by substituting $ m = 2 $ with $ m = -2 $ and recognizing that $ h_{2,-2} = (-1)^{2} h_{2,2}^{*} = h_{2,2}^{*} $. 
This simplification holds under the assumption that the progenitor spins are nearly aligned with the orbital angular momentum of the binary system~\cite{cotesta_analysis_2022}.

While the frequencies and damping time of the quasi-normal modes of a perturbed black hole depend only on its mass and spin, the amplitudes of the modes depend on how the black hole was excited. This implies that, for a black hole formed in the merger of a binary, the amplitudes depend on the progenitor parameters~\cite{kamaretsos_is_2012}. By fitting the amplitude dependence on progenitor parameters in numerical relativity simulations, a phenomenological expression for the amplitude $\mathcal{A}_{2,2}$ has been derived in~\cite{meidam_testing_2014}, 
\begin{equation}
     \mathcal{A}_{2,2} (\eta)=0.864 \eta \;,
\end{equation}
where $\eta=m_1 m_2 /\left(m_1+m_2\right)^2$ is the symmetric mass ratio, and $m_1$, $m_2$ are the masses of the progenitor black holes. Additionally, Berti et al.~\cite{berti_gravitational-wave_2006} derived fitting formulas  for the dimensionless mode frequencies $ f_{l,m}=M\omega_{l,m}/2\pi $, where $\omega_{l,m}$ are the angular frequencies of $(l,m)$ mode, and the quality factors $Q_{l,m}=\pi f_{2,2}\tau$, as functions of the dimensionless spin parameter $ j $. These fitting functions for the $(2,2)$ mode are accurate within $2\%$ of the numerical relativity results~\cite{berti_gravitational-wave_2006}.
\begin{subequations}\label{f_and_Q}
\begin{align}
& f_{2,2} \approx \frac{1}{2\pi}( 1.5251-1.1568(1-j)^{0.1292})\,,\\
& Q_{2,2} \approx 0.7000+1.4187(1-j)^{-0.4990}\,.
\end{align}
\end{subequations}
Therefore, the damping time 
\begin{equation}\label{eq:tau}
\tau \approx Q_{2,2}/(\pi f_{2,2}) \;. 
\end{equation}
The final mass $M$ and spin parameters $j$ of the remnant black hole can be determined from the mass and spin parameters of the progenitor binary, using the formula given in~\cite{tichy_final_2008,rezzolla_final_2008} that replicates the results of numerical relativistic simulations. For equal-mass, non-spinning binaries, $j$ is approximately equal to $0.69$, which is the value used in the following simulations. 

The GW-induced timing residual for the $I$-th pulsar in a PTA is given by 
\begin{eqnarray}\label{eq:res}
s^I(t; \lambda) & = & \int_0^t \mathrm{d} t^{\prime} \, z^I(t^{\prime}; \lambda)\;,\\
z^I(t; \lambda) & = & \frac{\nu^I(t; \lambda) - \nu_0^I}{\nu_0^I} \;,
\end{eqnarray}
where $\nu^I(t; \lambda)$ represents the observed spin frequency of the pulsar at the Solar System Barycenter (SSB), while $\nu_0^I$ is the intrinsic spin frequency at the pulsar. The signal is characterized by the set of parameters $\lambda$, which includes $\omega_{2,2}, \tau, t_0,  \varphi_{0}$ introduced earlier, right ascension $\alpha$, declination $\delta$, the inclination angle of the binary orbital angular momentum relative to the line of sight $\iota = \pi-\theta$, the overall amplitude $\zeta$, and GW polarization angle $\psi$. 
For a ringdown source, $z^I(t; \lambda)$ can be expressed as 
\begin{equation}\label{eq:doppler}
z^I(t ; \lambda)=\sum_{A=+, \times} F_A^I(\alpha, \delta) \Delta h_A\left(t ; \lambda_s\right)\;,
\end{equation}
where $\lambda_s = \{\omega, \tau, t_0, \varphi_{0}, \iota, \zeta, \psi\}$ with $\lambda = \{\alpha, \delta\} \cup \lambda_s$, and $F^{I}_{+,\times}$ are the antenna pattern functions for the $I$-th pulsar~\cite{2014ApJ...795...96W}. The term $\Delta h_A(t; \lambda_s)$ denotes the difference between the polarization waveforms at the SSB (Earth term) and the pulsar term, separated by the light travel time from the pulsar to the SSB. Given that the time delay between the Earth and pulsar term is several hundred years or more while the duration of a ringdown signal that exceeds the Nyquist frequency is at most a few weeks, an Earth term corresponding to the ringdown signal would be accompanied by a pulsar term that is the inspiral phase of the signal. Therefore, a search for the high-frequency ringdown signal would be insensitive to the significantly lower frequency pulsar term, and the latter may be safely ignored in the following. By combining Eq.~\ref{eq:waveform}, Eq.~\ref{eq:doppler}, and Eq.~\ref{eq:res}, we derive the Earth term timing residual $s_{e}^{I}$
\begin{widetext}
\begin{equation}\label{eq:residual}
s_e^I(t;\lambda) = \! \frac{Q_{2,2}^2}{2 (1+Q_{2,2}^2)\omega_{2,2}} \sum_{A=+,\times} F_{A}^{I} \sum_{m=\pm 2} h_{(l=2,m),A} + \frac{Q_{2,2}}{2 (1+Q_{2,2}^2)\omega_{2,2}} \sum_{m=\pm 2} \left[ F_{+} h_{(l=2,m),\times} - F_{\times} h_{(l=2,m),+} \right] + \mathcal{C}\;,
\end{equation}
\end{widetext}
In this expression, $\mathcal{C}$ is an integration constant ensuring that the timing residual is zero at $t = 0$. We focus on the oscillatory component of the signal, thereby omitting $\mathcal{C}$. The second term contributes marginally, accounting for only $\approx 15\%$ of the magnitude of the first term. 
Due to the low SNR of the ringdown signal, our analysis centers exclusively on the contribution from the primary term. 
In addition, we set $Q_{2,2}^2/(1+Q_{2,2}^2)\approx 1$, which is valid for high quality factors $Q_{2,2}$ typical for $j > 0.69$. Thus, the timing residuals for the Earth term, induced by $h_{+,\times}(t)$, can be approximated as 
\begin{equation}\label{eq:toa}
s_e^I(t;\lambda) = \frac{1}{2 \omega_{2,2}}\sum_{A=+,\times}F_{A}^{I}\sum_{m=\pm 2} h_{(l=2,m),A}\;. 
\end{equation}


\section{Detection and estimation of ringdown signals}\label{sec:Detect}

The ringdown signal detection problem addressed here consists of deciding between the following mutually exclusive hypotheses about the PTA data set comprised of the measured timing residuals $\mathbf{r}=\{\overline{r}^I\}$, $I = 1, 2, \ldots, N_p$, $\overline{r}^I_j = r^I(t_j)$, $j = 0,1,\ldots,N_I$. For a comprehensive discussion of hypothesis testing, see \cite{lehmann1986testing}.
\begin{enumerate}
    \item $\mathcal{H}_0$: The data consists of only noise, $ r^I(t) = n^I(t)$.
    \item $\mathcal{H}_\lambda$: The data contains a signal characterized by parameters  $\lambda$,  $ r^I(t) = n^I(t) + s^I(t; \lambda)$.
\end{enumerate}

For $\mathcal{H}_0$, the joint probability density function (PDF) of the data is given by
\begin{align}
    p(\mathbf{r} | \mathcal{H}_0) &= \prod_{I=1}^{N_p} p^I\left(\overline{r}^I\right) = \prod_{I=1}^{N_p} \frac{\exp\left(-\frac{1}{2} \langle \overline{r}^I | \overline{r}^I \rangle_I \right)}{\sqrt{(2\pi)^{N^I} |\boldsymbol{C}_I}|}\;,
    \label{eq:joint_p0}
\end{align}
where we assume the noise is Gaussian, and $\langle \overline{r}^I | \overline{r}^I \rangle_I$ is the noise weighted inner product for pulsar $I$, given by $\langle \mathbf{a} | \mathbf{b} \rangle_I = \mathbf{a}^T \boldsymbol{C}_I^{-1} \mathbf{b}$.  
Here, $\boldsymbol{C}_I$ is the covariance matrix of the noise process for the $I$-th pulsar, and $|\boldsymbol{C}_I|$ is its determinant. We assume that the noise processes across different pulsars are statistically independent at the high frequencies relevant to a ringdown search. 
For simplicity, even though our approach does not require it, we assume the noise in each pulsar to be white and stationary, which makes $\mathbf{C}_I$ diagonal with the non-zero elements being the variance $\sigma_I^2$. 
Under $\mathcal{H}_\lambda$, the joint PDF $p(\mathbf{r} | \mathcal{H}_\lambda)$ is obtained by replacing $\overline{r}^I$ with $\overline{r}^I - \overline{s}_\lambda^I$ in Eq.~\ref{eq:joint_p0}. 

In the binary hypotheses case, where $\lambda$ is known a priori, the optimal decision rule under the Neyman-Pearson criterion requires the computation of the log-likelihood ratio (LLR) given by 
\begin{eqnarray}
     \Lambda(\mathbf{r}) &=& \ln \frac{p(\mathbf{r} | \mathcal{H}_\lambda)}{p(\mathbf{r} | \mathcal{H}_0)} \;,
     \label{eq:LLR_def}
\end{eqnarray}
and comparison of $\text{LLR}$ with a threshold set according to a specified false alarm probability (deciding $\mathcal{H}_\lambda$ when $\mathcal{H}_0$ is true). For this case, the performance of the decision rule is completely quantified by the network SNR given by  $\rho_n^2(\lambda) = \sum_{I=1}^{N_p} \rho_{I}^2$, where 
\begin{equation}
  \label{snr}
\rho_{I}^{2}(\lambda) = \sum_{j=1}^{N_I} \frac{s^I(t; \lambda)^{2}}{\sigma_{I}^{2}} \;,
\end{equation}
is the SNR for the $I$-th pulsar. 

For the case of composite hypotheses where $\lambda$ is unknown, there is generally no optimal decision rule. In this case, the Generalized Likelihood Ratio Test (GLRT) provides a pragmatic solution by substituting the unknown deterministic parameters with their maximum likelihood estimates in the likelihood ratio. Thus, the decision rule compares the GLRT statistic given by
\begin{equation}
\text{GLRT}(\mathbf{r}) = \max_\lambda \ln \frac{p(\mathbf{r} | \mathcal{H}_\lambda)}{p(\mathbf{r} | \mathcal{H}_0)}   
\label{eq:glrt_def}
\end{equation}
with a threshold. 
\subsection{Separation of parameters} \label{sebsec:sep}
The maximization over $\lambda$ in Eq.~\ref{eq:glrt_def} is a challenging optimization problem given the multi-modality of the non-linear log-likelihood function $\Lambda(\mathbf{r})$ over the search space in $\lambda$ and its high dimensionality.
However, by splitting $\lambda$ into the subsets of extrinsic and intrinsic parameters, where the former can be maximized over more easily using analytic or computationally efficient techniques, the dimensionality of the search space for numerical optimization can be reduced. As shown below, the extrinsic parameters for the ringdown signal can be further reparametrized such that they appear linearly in the signal model and, hence, can be maximized over analytically (see Sec~\ref{sebsec:fs}). 

By combining Eq.~\ref{eq:waveform} and Eq.~\ref{eq:toa}, the Earth term timing residual for the $I$-th pulsar can be written as 
\begin{widetext}
\begin{equation}\label{eq:te}
\begin{aligned}
s_{e}^{I}(t;\lambda) &=\frac{\zeta}{\omega_{2,2}} e^{-\left(t-t_{0}\right) / \tau}\left[\left(F_{+}^{I} \cos 2 \psi-F_{\times}^{I} \sin 2 \psi\right)\left(1+\cos ^{2} \iota\right) \cos \left(\omega_{2,2} t+\varphi_{0}\right)\right.\\
&\left.+\left(F_{+}^{I} \sin 2 \psi+F_{\times}^{I} \cos 2 \psi\right) 2 \cos \iota \sin \left(\omega_{2,2} t+\varphi_{0}\right)\right]
\;.
\end{aligned}
\end{equation}
\end{widetext}
Following the approach used in the $\mathcal{F}$-statistic for continuous wave sources~\citep{1998PhRvD..58f3001J}, we can rewrite Eq.~\ref{eq:te} as follows 
\begin{equation}\label{eq:Fstat}
s^{I}_{e}(t ;\lambda) = \sum_{\mu=1}^{4}  a_{\mu}(\zeta,\iota,\varphi_{0},\psi) A_{\mu}^{I}(t;\alpha,\delta,\omega,\tau,t_{0}) \;, 
\end{equation}
where the coefficients $a_{\mu}$ are 
\begin{subequations}
\begin{align}
a_{1} &= X \cos2\psi \cos\varphi_{0} + Y \sin2\psi \sin\varphi_{0}    \;, \\
a_{2} &= -X \cos2\psi \sin\varphi_{0} + Y \sin2\psi \cos\varphi_{0}    \;, \\
a_{3} &= -X \sin2\psi \cos\varphi_{0} + Y \cos2\psi \sin\varphi_{0}    \;, \\
a_{4} &= X \sin2\psi \sin\varphi_{0} + Y \cos2\psi \cos\varphi_{0}    \;.
\end{align}\label{eq:ai}
\end{subequations}
In the equations above, $X$ is defined as $\zeta (1+\cos^2\iota)$ and $Y$ represents $2\zeta \cos\iota$. The time-dependent functions $A_{\mu}$ are given by 
\begin{subequations}
\begin{align}
A^I_1 + {\rm i} A^I_2 &= \varsigma F_+^I e^{{\rm i}\omega_{2,2} t} \;, \\
A^I_3 + {\rm i} A^I_4 &= \varsigma F_\times^I e^{{\rm i}\omega_{2,2} t} \;,
\end{align}\label{eq:Ai}
\end{subequations}
where $\varsigma =  e^{-(t-t_{0})/\tau}/\omega_{2,2}$ and $t_{0}$ is the starting time of the ringdown, and they depend only on the intrinsic parameters $\lambda_{i} = \lbrace \alpha,\delta,\omega,\tau,t_{0} \rbrace$. 

The set of coefficients ${a_{\mu}}$, $\mu=1,2,3,4$, are functions of the four extrinsic parameters $\lambda_{e} = \lbrace \zeta,\iota,\varphi_{0},\psi \rbrace$. For given ${a_{\mu}}$, we can solve for $\lambda_e$ as follows. Defining the variables,
\begin{subequations}
\begin{align}
A_{+} &= \sqrt{(a_{1}+a_{4})^2 + (a_{2}-a_{3})^2} \notag \\
&\quad + \sqrt{(a_{1}-a_{4})^2 + (a_{2}+a_{3})^2}\;,\label{eq:Ap} \\
A_{\times} &= \sqrt{(a_{1}+a_{4})^2 + (a_{2}-a_{3})^2} \notag \\
&\quad - \sqrt{(a_{1}-a_{4})^2 + (a_{2}+a_{3})^2}\;,\label{eq:Ac} \\
A &= A_{+} + \sqrt{A_{+}^2 - A_{\times}^2}\;, \label{eq:A}
\end{align}
\end{subequations}
we get~\cite{2012ApJ...756..175E,2007CQGra..24.5729C}
\begin{align}
\zeta &= \frac{A}{4} \;, \label{eq:zeta} \\
\iota &= \arccos\left(\frac{A_{\times}}{A} \right) \;, \label{eq:iota} \\
\psi &= \frac{1}{2}\arctan\left(-\frac{a_{4}A_{+}-a_{1}A_{\times}}{a_{2}A_{+}+a_{3}A_{\times}} \right) \;, \label{eq:psi} \\
\varphi_{0} &= \arctan\left(-\frac{a_{4}A_{+}-a_{1}A_{\times}}{a_{3}A_{+}+a_{2}A_{\times}} \right) + \pi \, \mathrm{H}\left(\sin(2\psi)\right) \;. \label{eq:phi_0}
\end{align}
Here $\mathrm{H}(x)$ is the unit step function. Finally, we transform $\varphi_{0}$ into the interval of $[0, \pi)$, which ensures that $\zeta$ is always positive. 

\subsection{\texorpdfstring{$\mathcal{F}$}{F}-statistic} \label{sebsec:fs}
Using Eq.~\ref{eq:joint_p0} and Eq.~\ref{eq:LLR_def}, we get~\citep{2014ApJ...795...96W,2015ApJ...815..125W} 
\begin{equation}\label{eq:loglambda}
\Lambda(\mathbf{r}) = \sum_{I=1}^{N_p}\langle r^I|s^I(\lambda)\rangle_I - \frac{1}{2}\sum_{I=1}^{N_p} \langle s^I(\lambda)|s^I(\lambda)\rangle_I \;.
\end{equation}
Inserting Eq.~\ref{eq:Fstat} into Eq.~\ref{eq:loglambda}, we obtain 
\begin{equation}\label{eq:lognetwork}
\Lambda(\mathbf{r}) = \sum_{\mu=1}^{4}a_{\mu} N_{\mu}-\frac{1}{2}
\sum_{\mu=1}^{4}\sum_{\nu=1}^{4}a_\mu a_\nu M_{\mu \nu} \;,
\end{equation}
where $N_{\mu}=\sum_{I=1}^{N_p} \langle r^I|A_\mu^I \rangle_I$ is a $4\times 1$ vector that contains the data and the intrinsic parameters, $M_{\mu\nu}=\sum_{I=1}^{N_p}\langle A_\mu^I|A_\nu^I\rangle_I$ is a $4\times4$ matrix that contains only the intrinsic parameters. 
Maximizing $\Lambda(\mathbf{r})$ over the extrinsic parameters $\lambda_e$, or equivalently ${a_\mu}$, $\mu=1,2,3,4$, we get the $\mathcal{F}$-statistic,
\begin{equation}\label{eq:maxloglike2}
\mathcal{F}(\lambda_i)=\max_{\{a_{\mu}\}} \lbrace \Lambda(\mathbf{r})\rbrace = \frac{1}{2} N_{\mu} M^{\mu\nu} N_{\nu}  \;, 
\end{equation}
where $a_{\mu}= M^{\mu\nu} N_{\nu}$ is the value at which the maximum is obtained. 
To obtain the GLRT, one must search the intrinsic parameter space to find the maximum of $\mathcal{F}(\lambda_i)$. The location of the maximum provides the estimated values of the intrinsic parameters, which are then used to determine $a_{\mu}$ and, using the formalism given in Sec.~\ref{sebsec:sep}, the extrinsic parameters $\lambda_e$.

\subsection{Particle Swarm Optimization} \label{sebsec:pso}
For the optimization of the $\mathcal{F}$-statistic over the intrinsic parameter space, we employ PSO. This algorithm has been widely utilized across several applications in gravitational wave data analysis \citep{wang_particle_2010, weerathunga_performance_2017, PhysRevD.106.023016,zhang_resolving_2021}. 
In the following section, we present an overview of the PSO algorithm and list the specific parameter settings used in our analysis. 

PSO is an iterative stochastic method for finding the global maximum of a function $f(x)$, $x \in \mathbb{R}^N$, often referred to as the fitness function, over a compact subset $D \subset \mathbb{R}^N$ known as the search space.
In the context of our work, the fitness function corresponds to the $\mathcal{F}$-statistic (c.f., Eq.~\ref{eq:maxloglike2}), and the search space is the intrinsic parameter space.
Through each iteration, the method samples the fitness function at a set of locations, using these samples to update the locations for subsequent iterations. 
These sampling locations, referred to as particles, generally remain constant in number throughout the process, and the group of particles is known as a swarm. 
The positional update of each particle uses a vector, called its velocity, which is also updated iteratively.

Denoting the position and velocity of each particle at an iteration step $k$ by $x_{i,j}[k]$ and $v_{i,j}[k]$, respectively, where $i \in \{1, 2, \ldots, N_{\rm part}\}$ is the particle index and $j$ is the component index, the update rules are stated as follows.
\begin{subequations}
\begin{align}
x_{i,j}[k+1] &= x_{i,j}[k] + \min(v_{i,j}[k+1], v_{\rm max})\;, \\
v_{i,j}[k+1] &= w[k] v_{i,j}[k] + c_1 r_1 (p_{i,j}[k] - x_{i,j}[k]) \notag \\
&\quad + c_2 r_2 (l_{i,j}[k] - x_{i,j}[k])\;,
\end{align}
\end{subequations}
where $v_{\rm max}$ is a cap on the maximum step size in any direction, $w[k]$ is the inertia weight that decreases deterministically over iterations, and $c_1$, $c_2$ are called acceleration constants. The random variables $r_1$ and $r_2$ are uniformly distributed within the interval $[0, 1]$.
The term $p_i[k]$ represents the personal best position for particle $i$ based on its fitness history, while $l_i[k]$ signifies the best position among its local neighborhood. 
If the neighborhood includes all particles, $l_i[k]$ equates to the global best $g[k]$.
For neighborhoods, a common structure is the ring topology, where particle indices form a circle, and subsets of consecutive indices define a neighborhood.
 
Each term in the velocity update serves a distinct purpose: the inertia term propels the particle beyond its current position, aiding in the avoidance of local maxima and fostering search space exploration; the cognitive term pulls the particle towards the best solution in its history; the social term pulls it to the best solution identified in the history of its neighborhood.
While the cognitive and social terms encourage investigation of promising areas, their randomness impedes premature convergence. Thus, a balance is struck between exploration and exploitation, with the inertia weight  $w[k]$  controlling this trade-off by decreasing as iterations progress. Opting for a local rather than a global best can further sustain the exploration phase by curbing rapid information spread within the swarm.

PSO tends to excel when $D$, the search space, is hypercubic, constraining each $x$ component to an interval $[a_j, b_j]$.
Typically, the process begins with random initial positions and velocities within these ranges, ensuring the initial updates remain within $D$. However, to address potential exits from $D$ in later iterations, various boundary conditions can be applied.
Here, we utilize the ``let-them-fly" condition, where particles outside $D$ receive a fitness of $-\infty$. As for the termination of iterations, we adhere to an iteration count limit, with the final solution being the global best location and its associated fitness.

Despite the lack of guarantee for convergence to the global maximum, PSO can be adjusted to achieve a satisfactory probability, $P_{\rm success}$, of successful convergence to an optimal region. By executing $N_{\rm runs}$ independent PSO iterations, each with unique pseudo-random sequences, and selecting the best outcome from these, the success probability can be exponentially boosted to $1 - (1 - P_{\rm success})^{N_{\rm runs}}$.
Here we have opted to run $8$ sets of PSO in parallel on the data to increase the probability of converging to the global optimum. 

Most of the strategies discussed are broadly applicable to gravitational wave data analysis using PSO. 
Our parameter settings largely mirror those of previous work, including particle count $N_{\text{{part}}} = 40$ and iterations $N_{\text{iter}} = 2000$, which facilitates thorough search space exploration. We use $c_1 = c_2 = 2.0$ and $m = 3$ for neighborhood size. We set the maximum velocity $v_{\max} =  (b-a)/2$ at initialization and $ (b-a)/5$ for subsequent iterations. The inertia weight $w(k) = 0.9 - 0.5(k/(N_{\text{{iter}}}-1))$ decays linearly to balance exploration and exploitation.
\section{Results}\label{sec:SimResults}
Our main results characterizing the performance of the ringdown search method are presented in this section. We start with a description of our simulation setup, followed by discussions of the detection and parameter estimation performance of the method.
\subsection{Simulation} \label{sebsec:sim}
We generated a simulated timing residual data set by selecting the $100$ nearest pulsars from a synthetic catalog \cite{smits2009pulsar} of MSPs generated for the SKA. Each data set consists of independent white Gaussian noise realizations for each pulsar added to its corresponding (Earth term) GW ringdown signal.
The latter is first generated on a dense and uniformly spaced grid of time values, followed by retaining the samples closest to the set of observations times $\{t_j^I\}$, $j = 1, 2, \ldots, N_I$, for the $I$-th pulsar.

Asynchronous sampling times are generated by initially generating them for the $I$-th pulsar as $ t_j^I = (j-1)/(2f_{\rm sp}) + (I-1)/(2N_p f_{\rm sp}) $ with the sampling interval $1/(2f_{\rm sp})$ set to $2$ weeks. 
This scheme implies a constant time shift between the samples of different pulsars, ensuring that the samples for each pulsar are uniformly spaced. 
The duration of the simulated observation is $5$ years, resulting in $N_I=130$. Next, we employ the scheme from \cite{wang_extending_2021}, in which a random number $c_j^I$ drawn from a truncated Cauchy probability density function (PDF)~\cite{papoulis1991random} is added to $t_j^I$. 
The location parameter and the scale factor of the Cauchy PDF are set to zero and $1/3$~day, respectively, and the absolute value of $c_j^I$ is restricted to $\leq 7$ days. 
This randomized staggered sampling reflects a more realistic situation in scheduling astronomical observations. 
The truncated Cauchy distribution ensures that for a single pulsar, the consecutive sampling times \(t_j^I\) and \(t_{j+1}^I\) always satisfy \(t_{j+1}^I > t_j^I\). 

We simulated ringdown signals (cf., Sec.~\ref{sec:Ringdown waveform}) from a binary progenitor consisting of equal-mass non-spinning supermassive black holes.
The intrinsic parameters of the signals were sampled from a wide range as listed in Table~\ref{tab:parameter}.  
In all cases, the spin parameter was $j=0.69$, which leads to different damping times for the signals as shown in Table~\ref{tab:parameter}. 
The timing residual induced by a given GW source was calculated for each pulsar in the PTA using Eq.~\ref{eq:Fstat} and the sampling times $\{t^I_j\}$, $j = 1, 2, \ldots, N_I$. A realization of the full PTA data set was then obtained by adding an independently drawn random value from a zero mean Gaussian distribution with a standard deviation of $\sigma_n =10^{-7}$~sec to each sample. 

For statistical characterization of the detection performance of the method, five different values of the network SNR, $\rho_n \in \{0, 7, 10, 20, 100\}$, were used for the source with chirp mass $M_{c1}$ and position $P_1$, and $500$ independent PTA data realizations were generated for each $\rho_n$. The data realizations for the noise-only case ($\rho_n = 0$) were used to estimate the distribution of the fitness values under the null hypothesis $\mathcal{H}_0$. 
To reduce the computational burden of our analysis, we only use $\rho_n = 10$ and 100 PTA data realizations for each of the remaining sources in Table~\ref{tab:parameter}. 
\begin{table*}[htbp]
    \centering
    \caption{Parameters of the simulated ringdown signals for six sources and their search ranges. The angular frequency is expressed as a multiple of $\omega_{\rm sp}$, where $\omega_{\rm sp}=2\pi f_{\rm sp} = 81.96$ rad/yr. }
    \begin{ruledtabular}
    \begin{tabular}{lccccccc}
    \textbf{Parameter} & \textbf{Source 1} & \textbf{Source 2} & \textbf{Source 3} & \textbf{Source 4} & \textbf{Source 5} & \textbf{Source 6} & \textbf{Range} \\ \hline
    $\omega/\omega_{\rm sp}$ & 2 & 6 & 2 & 6 & 2 & 6 & $\left[0.0244,10\right]$ \\
    $\tau$/yr & 0.0396 & 0.0132  & 0.0396 & 0.0132 & 0.0396 & 0.0132 & $\left[7.9187\times 10^{-4},3.2451\right]$ \\
    $M_c/M_\odot$ & $M_{c1}$ & $M_{c2}$ & $M_{c1}$ & $M_{c2}$ & $M_{c1}$ & $M_{c2}$ & -- \\
    & $9.52\times 10^{9}$ & $3.17\times 10^{9}$ & $9.52\times 10^{9}$ & $3.17\times 10^{9}$ & $9.52\times 10^{9}$ & $3.17\times 10^{9}$ & -- \\ $(\alpha,\delta)$/rad & \multicolumn{2}{c}{$P_1(1.985, 0.625)$} & \multicolumn{2}{c}{$P_2(3.500, 0.300)$} & \multicolumn{2}{c}{$P_3(4.367, 0.880)$} & $\alpha: [0,2\pi]$, $\delta: [-\frac{\pi}{2},\frac{\pi}{2}]$ \\
    $\varphi_{0}$/rad & 0.8 & 0.8 & 0.8 & 0.8 & 0.8 & 0.8 & -- \\
    $t_0$/yr & 1.5 & 1.5 & 1.5 & 1.5 & 1.5 & 1.5 & $\left[0, 5\right]$ \\
    $\psi$/rad & 0.5 & 0.5 & 0.5 & 0.5 & 0.5 & 0.5 & -- \\
    $\iota$/rad & 0.4949 & 0.4949 & 0.4949 & 0.4949 & 0.4949 & 0.4949 & -- \\
    \end{tabular}
 \end{ruledtabular}\label{tab:parameter}
\end{table*}

\begin{figure}[htbp]
    \centering  \includegraphics[scale=0.6]{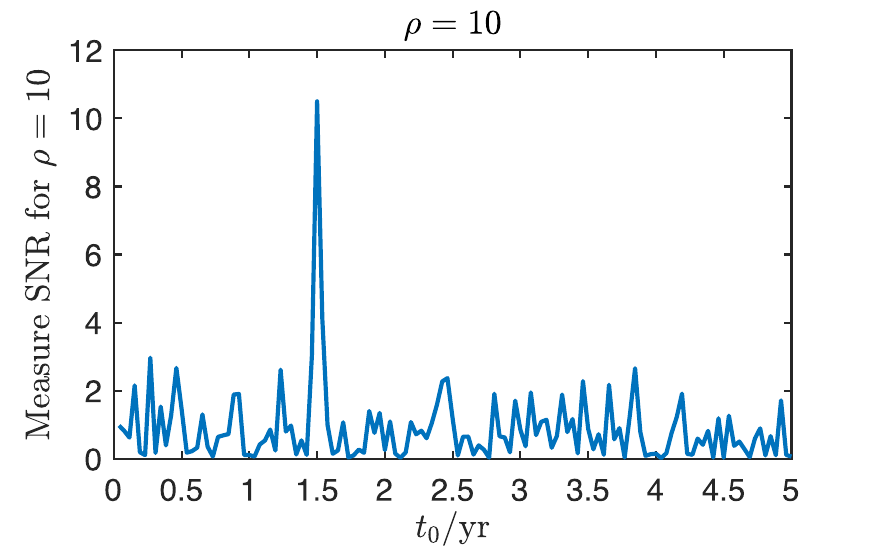}
    \caption{The measured SNR $\rho_m(\lambda)$ as a function of start time $t_0$ for a network SNR $\rho_n=10$ injected signal with a true start time of $1.5$~yr. The SNR measured at $t_0=1.5$~ year is $10.50$, statistically consistent with the injected value. }       
    \label{measuresnr}
\end{figure}

To verify the correctness of the amplitude normalization for injected signals, we compare the measured SNR with that of the injected signal. The measured SNR $\rho_m(\lambda)$ is defined as 
\begin{equation} \label{measnr}
\rho_m(\lambda) = 2 \times \left(\frac{ |\mathcal{F} - E[\Lambda|\mathcal{H}_0]|}{\sqrt{\operatorname{Var}[\Lambda|\mathcal{H}_0]}}\right)^{\frac{1}{2}}\;,
\end{equation}
where $E[\Lambda|\mathcal{H}_0]$ and $\operatorname{Var}[\Lambda|\mathcal{H}_0]$ represent the expectation and variance of the fitness under the noise-only hypothesis $\mathcal{H}_0$, respectively. If the normalization of an injected signal is correct, $\rho_m(\lambda)$ should be approximately normally distributed with the mean equal to the injected SNR when $\lambda$ matches the true injection parameters. Figure~\ref{measuresnr} shows $\rho_m(\lambda)$ where all the intrinsic parameters were set to be equal to the injected ones except for the start time of the signal $t_0$ with the true injected start time and network SNR being $1.5$ years and $10$, respectively. 

Table~\ref{tab:parameter} also shows the search range settings for intrinsic parameters. 
For the simulated PTA described above, $f_{\rm PTA}=N_pf_{\rm sp} \approx 8196$ rad/yr is the highest theoretically detectable frequency under staggered sampling. It should be noted, however, that unlike continuous signals, ringdown signals have a finite duration. When the duration is comparable to or shorter than the average sampling interval, some pulsars may not contribute to the detection of the signal because their observation times may not fall within the signal. Since, in our case, the duration of the ringdown signal decreases with its frequency (cf., Eq.~\ref{eq:tau}), this implies an upper bound on the signal frequency that is less than $f_{\rm PTA}$.
Assuming that only the pulsars observed within one damping time of a ringdown signal with frequency $f_h$ can contribute to its detection, we get 
\begin{equation}
    f_h=\frac{\tau}{\Delta T} N_p f_{\rm sp} = \frac{Q_{2,2}}{\pi f_h}\frac{f_{\rm sp}}{2}N_p f_{\rm sp} \;.
\end{equation}
Therefore, the highest detectable frequency of the ringdown signal under the current simulation setup and signal model is $\sqrt{N_p Q_{2,2}/2\pi}f_{\rm sp}< N_P f_{\rm sp}$.
For the lower frequency limit of our search, we use the realistic upper limit~\cite{berti_gravitational-wave_2006} of $7.8\times 10^{11} M_{\odot}$ on the progenitor chirp mass, which leads to $2$~rad/yr as the lower limit. 

Given that frequency and damping time can span several orders of magnitude, it is best to initialize the PSO particles with a log-uniform distribution along the frequency and damping time parameters instead of a uniform distribution. This ensures that there is no deficit of particles at low values of frequency and damping time, which could reduce the effectiveness of the PSO in searching for the global maximum of the fitness function. For the remaining intrinsic parameters $\alpha$, $\delta$, $t_0$, the initialization is done using a uniform distribution. 
\subsection{Detection performance} \label{sebsec:det}
\begin{figure}[htbp]
    \centering
\includegraphics[scale=0.9]{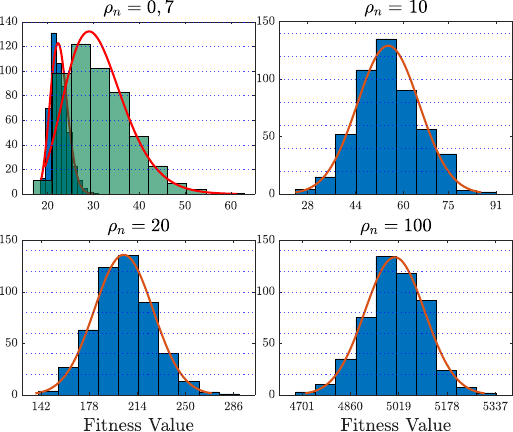}
    \caption{
    The distribution of the fitness value for 500 realizations with $\rho_n=0$, $7$, $10$, $20$, and $100$. The red curve in each panel shows the best-fit distribution. These are $\ln \mathcal{N}(\mu=3.11, \sigma= 0.08)$ , $\ln \mathcal{N}(\mu=3.41, \sigma= 0.21)$, $\mathcal{N}(\mu=54.94, \sigma=10.36)$, $\mathcal{N}(\mu=203.84, \sigma=22.02)$, and $\mathcal{N}(\mu=5006.32, \sigma=98.62)$, respectively. }
    \label{fitness}
\end{figure}

The distribution of the fitness for various SNR values is illustrated in Fig.~\ref{fitness}. From the distribution of the fitness under $\mathcal{H}_0$, we get a false alarm probability of $\lesssim 0.2\%$ if the detection threshold is set at the maximum observed fitness value of $30.1$. 
We experimented with some low SNR values and found that a detection probability of $\approx 50\%$ is achieved when $\rho_n =7$. 
The distribution of the fitness value evolves into a nearly normal distribution for higher SNR values, as shown in
Fig.~\ref{fitness} for $\rho_n=10$, $20$ and $100$. For $\rho_n\gtrsim 20$, the measured detection probability reaches unity. 
In a system with a chirp mass of $9.52 \times 10^{9} M_\odot$, SNR values of $7$ and $20$ correspond to distances of approximately $600$ Mpc and $210$ Mpc, respectively.

\begin{figure}[htbp]
    \centering
   \includegraphics[scale=0.45]{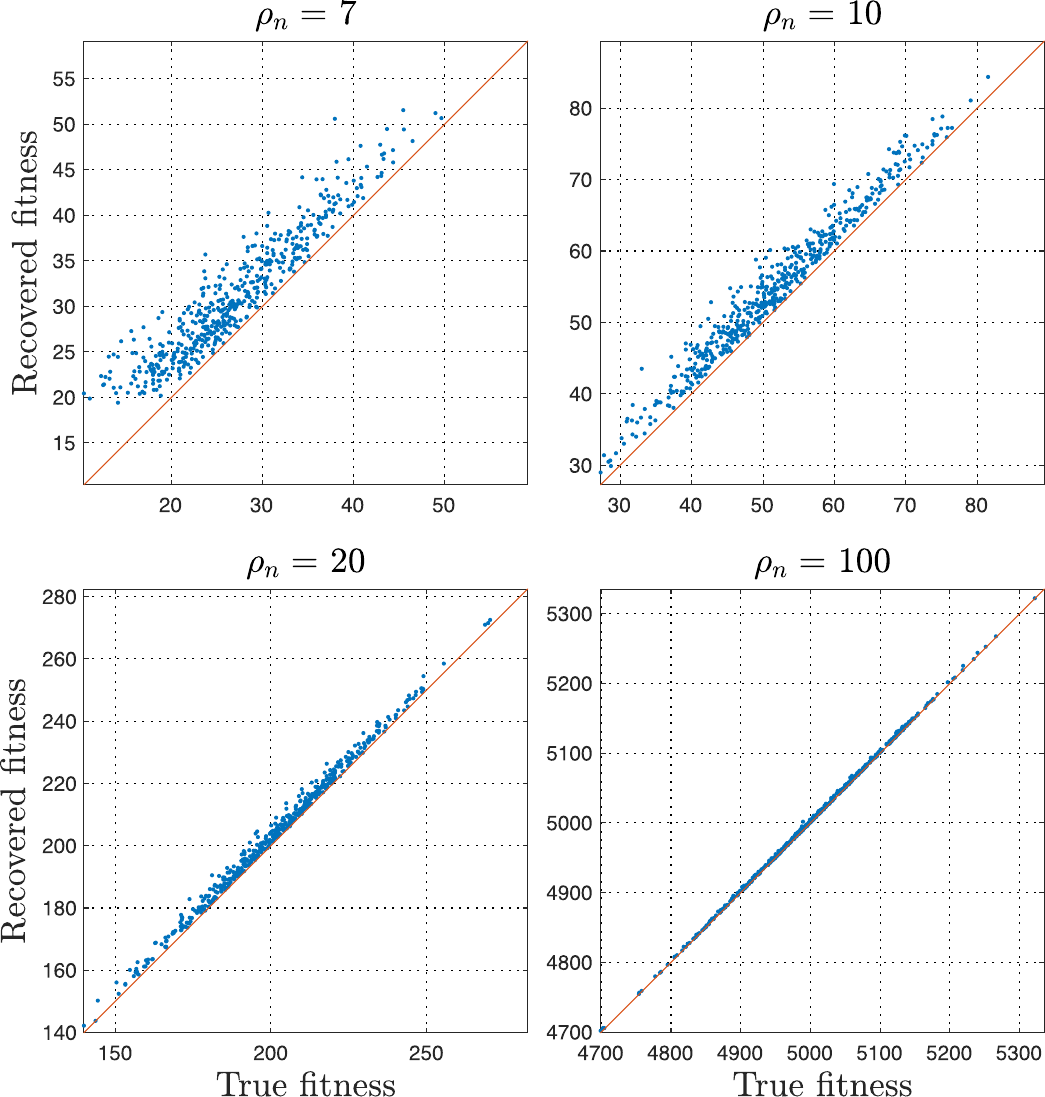}
    \caption{
    Recovered fitness values from the best of eight PSO runs versus the true fitness values from the injected parameters for 500 realizations with $\rho_n$ = 7, 10, 20, and 100.}
    \label{lr}
\end{figure}

In general, practical stochastic global optimization algorithms like PSO do not have a guarantee of convergence to the global maximum (even asymptotically). However, we can gauge the effectiveness of PSO on a given data analysis problem by simulating multiple data realizations and checking, for each realization, if the fitness returned by PSO exceeds the one at the injected signal parameters. If this happens in the majority of realizations, our confidence in the performance of PSO for the given data analysis problem is affirmed. 
Figure~\ref{lr} exhibits the best fitness value obtained from eight PSO runs versus the fitness from the injected parameters with $\rho_n=0$, $7$, $10$, $20$, and $100$. It can be seen that PSO found a better fitness value than the one at the injected parameters in all simulated data realizations, indicating that the lack of convergence to the global maximum is not a significant issue for our analysis. 
\subsection{Estimation performance} 
Table~\ref{para_9} presents the bias and standard deviation of intrinsic parameter estimates for each source. 
\begin{table*}[htbp]
    \centering
    \caption{Biases and standard deviations of the intrinsic parameters $\alpha, \delta, \omega, \tau, t_0$ calculated for the sources listed in Table~\ref{tab:parameter}, based on $500$ realizations for Source $1$ and $100$ realizations for the other sources. The SNR of all sources is $10$. The bias is defined as the deviation of the mean from the true value. Each parenthesis contains the bias and standard deviation in the format (Bias, Std). The units are consistent with those in Table~\ref{tab:parameter}.}
    \begin{ruledtabular}
    \begin{tabular}{lccccc}
    \textbf{Source} & $\boldsymbol{\alpha}$ & $\boldsymbol{\delta}$ & $\boldsymbol{\omega}$ & $\boldsymbol{\tau}$ & $\boldsymbol{t_0}$ \\ \hline
    Source 1 & $(-0.0024, 0.0865)$ & $(-0.0144, 0.0677)$ & $(0.0017, 0.0877)$ & $(0.0026, 0.0109)$ & $(-0.0015, 0.0021)$ \\
Source 2 & $(-0.0318, 0.1028)$ & $(0.0260, 0.0766)$ & $(-0.0552, 0.6431)$ & $(0.0004, 0.0039)$ & $(-0.0004, 0.0006)$ \\
Source 3 & $(-0.0151, 0.1071)$ & $(-0.0095, 0.0553)$ & $(0.0017, 0.0730)$ & $(0.0024, 0.0110)$ & $(0.0001, 0.0025)$ \\
Source 4 & $(-0.0296, 0.1596)$ & $(-0.0412, 0.0948)$ & $(0.0422, 0.2702)$ & $(0.0011, 0.0037)$ & $(-0.0002, 0.0010)$ \\
Source 5 & $(0.0001, 0.1218)$ & $(0.0080, 0.0491)$ & $(0.0000, 0.0822)$ & $(0.0020, 0.0104)$ & $(0.0000, 0.0021)$ \\
Source 6 & $(-0.0513, 0.4140)$ & $(-0.0205, 0.2099)$ & $(0.0105, 0.2242)$ & $(0.0013, 0.0034)$ & $(-0.0002, 0.0011)$ \\
    \end{tabular}
    \end{ruledtabular}\label{para_9}
\end{table*}
We see that the bias in all cases is within the standard deviation, which means that the algorithm has negligible bias over a wide range of SNR and intrinsic parameters. In general, the bias and standard deviations are seen to increase as the angular frequency becomes larger and decay times become shorter. 
The sky location parameters $(\alpha, \delta)$ generally show good agreement with the true values, although some sources, like Source $6$, exhibit larger standard deviations, indicating greater positional uncertainty for certain signals. 
Notably, the signal start time $t_0$ estimates are remarkably precise for all sources, with mean values nearly identical to the true value of $1.5$ years and small standard deviations. Accurate signal start time estimates are extremely beneficial for archival surveys in multi-messenger astronomy, such as the LSST, as it can narrow the search window and help identify the corresponding objects~\cite{2017arXiv170804058L, Bianco_2022}.

\begin{figure}[htbp]
    \centering
\includegraphics[scale=0.85]{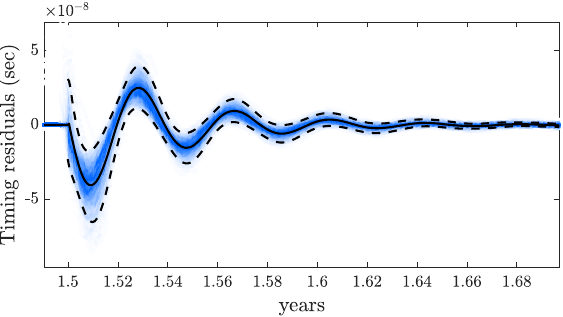}
     \caption{
    The black line depicts the injected signal for one of the pulsars in the PTA with an SNR of $10$. The blue shading indicates the reconstructed signals from 500 data realizations. The area enclosed by the black dashed line represents the $90\%$ confidence region for the waveform reconstruction. This confidence region was determined by calculating the mean of the reconstructed waveform at each time point and adding and subtracting 1.96 standard deviations, thereby capturing the range within which the reconstructed waveform is expected to lie with $90\%$ probability.
    }
    \label{res 10}
\end{figure}

Taking Source $1$ as an example, we examine an SMBBH characterized by a chirp mass of $9.52 \times 10^{9} M_\odot$ situated at a distance of approximately $420$ Mpc. The signal frequency for this source is set at twice the Nyquist frequency. 
Figure~\ref{res 10} exhibits the reconstruction of the ringdown signal waveform. On average, the start time of the reconstructed signal is earlier by approximately 13 hours, indicating a systematic temporal bias. 
The curve formed by averaging all the estimated signals at each time point closely matches the injected signal. This indicates that the algorithm performs well in terms of signal shape estimation, even though there is a minor start-time discrepancy.

\begin{figure*}
    \centering
\includegraphics[scale=0.65]{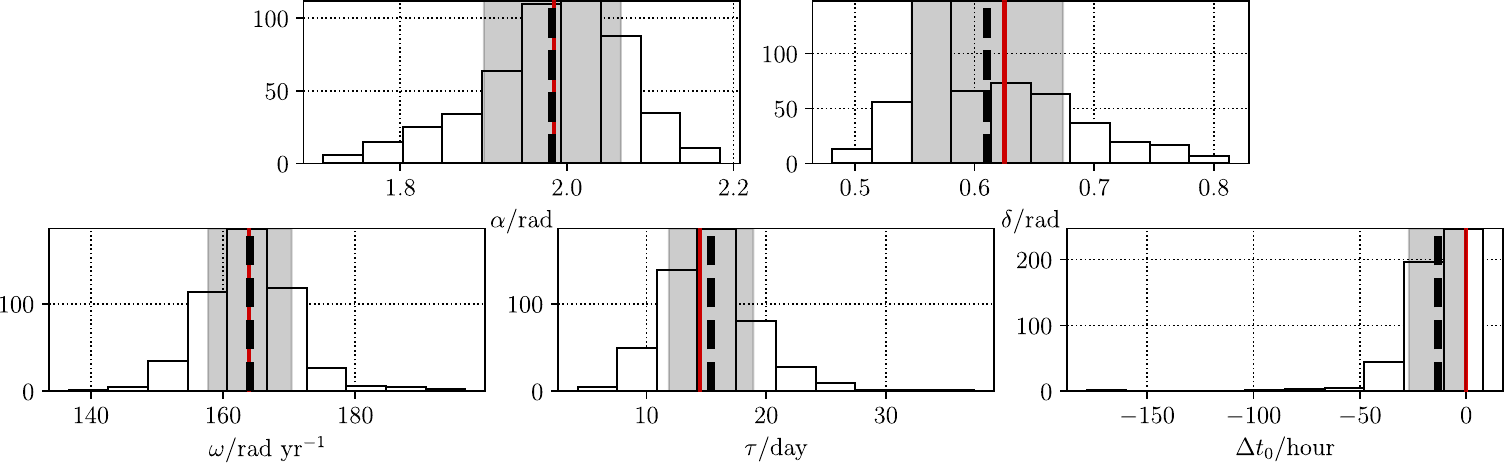}
    \caption{Histograms of the estimated intrinsic parameters of $500$ data realizations for Source $1$. The red vertical solid line represents the true value of the parameters used in the simulation, the black vertical dashed line represents the mean value, and the gray shaded area covers the $\pm 1$ standard deviation interval centered on the sample mean. }
    \label{para10}
\end{figure*}

Figure~\ref{para10} illustrates the distribution of the estimated intrinsic parameters $\{\alpha,\delta,\omega,\Delta t_0,\tau\}$ at a SNR of $10$. Here, $\Delta t_0$ represents the estimation error of the signal start time $t_0$. The injected simulation parameters are indicated by red vertical solid lines, while the estimated mean values are represented by black vertical dashed lines. 
The results suggest that accurate estimation of the Right Ascension, declination, and frequency of GW is achievable, with only small biases observed across all parameters. 
However, as mentioned earlier in connection with Fig.~\ref{res 10}, the estimation of the signal start time $t_0$ exhibits a systematic bias, with the estimated value typically earlier by approximately $13$ hours relative to the true value. 
Despite this delay, the estimation may still be sufficiently accurate for follow-up electromagnetic (EM) observations, given the LSST's observing strategy. This strategy involves repeated observations of the same point in the sky, typically spaced from a few days to several weeks apart~\cite{2017arXiv170804058L}.
The one-sigma contour of the source location distribution has an enclosed area of approximately $17.33~\text{deg}^2$ on the sky, with $\sigma_\alpha$ approximately equal to $4.97$ degrees and $\sigma_\delta$ approximately equal to $3.88$ degrees. This area is comparable to the LSST’s $9.6$ square degree field of view, meaning that a single LSST pointing can effectively cover most of the error region. 
Furthermore, the frequency of GW can be accurately estimated with a standard deviation of $7.20 \, \text{rad} \, \text{yr}^{-1}$. These precise localization and frequency measurements enhance the capability to correlate gravitational wave signals with electromagnetic counterparts. 
Combining the estimation results of the parameters $\omega$ and $\tau$, we obtain an estimated mean of $3.46$ for the quality factors $Q$, with an estimated variance of $0.90$ and a relative error of $6.51\%$. 
Combining with Eq.~\ref{f_and_Q}, we estimate the spin parameter $j$ and final total mass $M$ of the binary black hole progenitor to be $0.64$ and $2.10 \times 10^{10} M_\odot$, respectively, with a bias of $0.05$ and $2.09 \times 10^{8} M_\odot$, and variances of $0.27$ and $3.21 \times 10^{9} M_\odot$. The estimated parameters for all sources are presented in Table~\ref{para_3}. 
\begin{table*}[htbp]
    \centering
    \caption{Bias and standard deviation of the parameters $Q$, $j$, $M$ based on $500$ realizations from source $1$ and $100$ realizations from other sources. Bias and standard deviation of parameters $Q$, $j$, and Bias \% and Std \% of mass parameter $M$ relative to the true mass. Each parenthesis contains the bias and standard deviation in the format (Bias, Std).}
    \begin{ruledtabular}
    \begin{tabular}{lcccc}
    \textbf{Source} & \boldmath$Q$ & \boldmath$j$ & \boldmath$M$ Bias (\%) & \boldmath$M$ Std (\%) \\ \hline
    Source 1 & $(0.2113, 0.8954)$ & $(-0.0477, 0.2690)$ & 0.957\% & 15.43\% \\
    Source 2 & $(0.1023, 0.9858)$ & $(-0.0548, 0.2030)$ & -0.435\% & 14.57\% \\
    Source 3 & $(0.1990, 0.9272)$ & $(-0.0462, 0.2369)$ & 0.561\% & 15.01\% \\
    Source 4 & $(0.3173, 0.9913)$ & $(-0.0391, 0.2760)$ & 2.31\% & 16.68\% \\
    Source 5 & $(0.1623, 0.8564)$ & $(-0.0399, 0.1996)$ & 0.325\% & 13.86\% \\
    Source 6 & $(0.3369, 0.8335)$ & $(0.0087, 0.1805)$ & 3.78\% & 12.88\% \\
    \end{tabular}
    \end{ruledtabular}
    \label{para_3}
\end{table*}

Combined with observations during the inspiral and merger phases, a precise estimate of the total mass will allow us to constrain physical parameters of the binary black hole system such as the mass ratio. 
The estimated mass and spin parameters are crucial for understanding the formation and evolution of supermassive binary black holes. The spin size is closely related to the formation history of the black hole. A high spin value may mean that the black hole has experienced a long period of accretion or multiple merger events~\cite{berti_quasinormal_2009}. 

Compared to space-based gravitational wave detectors, for sources with mass $M > 10^7 M_\odot$, the ability of space-based detectors to estimate spin and mass using ringdown signals diminishes as the mass increases, whereas the detection capability of PTA correspondingly strengthens. For sources with a mass ratio of 2, a residual mass of approximately $10^9 M_\odot$, and a redshift $z < 0.7$, LISA can constrain mass and spin within $10\%$~\cite{baibhav_lisa_2020}. This is comparable to the mass estimation results obtained from our simulated PTA. Considering the anticipated further increase in the number of pulsars in future PTA and their significantly better timing precision, constraints on the parameters of supermassive binary black holes using ringdown signals may reach or even surpass those of space-based gravitational wave detectors.

\section{Conclusions}\label{Sec:conc}
We present a coherent network analysis method, utilizing staggered sampling and GLRT, for detecting and estimating ringdown signals using pulsar timing arrays. Specifically, our method divides the parameters of the ringdown signal into intrinsic and extrinsic parameters, considering only the $(2, 2)$ mode. Extrinsic parameters are determined analytically, while intrinsic parameters are numerically determined using PSO.

We developed a simulated data set spanning five years, comprising 100 pulsars and only considering white noise.
We simulated six SMBBH sources with different masses and spatial positions.
Using the first wave source, we evaluated the detection capability of this method at SNRs of $7$, $10$, $20$, and $100$.
The algorithm demonstrates good usability in different SNR scenarios, which is reflected in the fact that in all $500$ realizations, the fitness values obtained by the algorithm are higher than those corresponding to the real parameters.
Furthermore, we used all six sources to evaluate the effectiveness of this method in estimating ringdown signal parameters. 
Our findings suggest that under optimal conditions, the PTA can detect and accurately retrieve ringdown signals with a frequency of $f=2f_{\rm sp}$. 
For ringdown signals with higher frequencies, unlike continuous source signals, the parameter estimation capability decreases due to the shortened time window of the ringdown signal, which results in fewer pulsars within the window.

The reported results were obtained subject to certain limitations. It is assumed that all pulsars exhibit stationary white Gaussian noise with zero mean and the same variance. In reality, pulsar timing residuals can exhibit more complex noise characteristics, including red noise and non-stationary behavior. While Finn~\cite{finn_aperture_2001} has demonstrated that coherent methods like ours are robust against non-Gaussianity in timing residual noise, the uniform noise variance assumption may not hold for real data sets. 
Future work should aim to incorporate more sophisticated noise models that account for varying noise levels and non-stationary characteristics across different pulsars. 
The noise covariance matrix including different red noise components should be incorporated into the algorithm after evaluating the red noise for each pulsar. 
Alternatively, the red noise can be suppressed before the search for ringdown signals by employing adaptive spline fitting techniques. These methods involve modeling the low-frequency trend of red noise using B-spline functions with knot placement optimized via particle swarm optimization, followed by subtraction of the fitted noise to enhance the signal-to-noise ratio~\cite{mohanty_adaptive_2021,mohanty_glitch_2023}. Although not yet applied to PTA data, ongoing research efforts are focused on adapting this approach to suppress the red noise in individual pulsars~\cite{qian_novel_2024}.
In addition, timing models play a critical role in GW searches. We need to ensure that the parameters, especially the dispersion measures, in the timing models are properly taken into account.
The early PTA data were observed highly uneven, and the noise level varies over the years; therefore, only a limited subset of pulsars has been adopted in the search for deterministic signals~\cite{antoniadis_second_2024}. The selection of pulsars is indeed an indispensable part of the actual data processing, especially for the ringdown signals, since we need to strike a balance between the data quality and the number of pulsars, the latter determining the highest frequency we can reach. All of these are important steps of our method towards an implementation to the real PTA data sets.

Our analysis exclusively considers the $(2,2)$ mode in the ringdown waveform of gravitational waves. Higher-order modes, which can provide additional information about the source's properties, were deliberately excluded due to the current focus on optimizing SNR requirements. Resolving these higher-order modes necessitates a relatively high SNR, a goal that extends beyond the scope of the present study. However, incorporating higher-order modes can significantly enhance the analysis by providing more detailed insights into the mass ratio of the progenitor and spin of the final black hole, improving the understanding of the geometric configuration of the source, and enabling more stringent tests of general relativity. We anticipate that including higher-order modes would not significantly alter our primary outcomes but acknowledge that their incorporation could enhance the depth and accuracy of parameter estimations. Future investigations should explore methods to effectively include these additional modes, especially as data quality and processing techniques improve.

Currently, our study focuses solely on the ringdown phase of gravitational wave signals. However, an SMBBH signal has multiple phases, including the inspiral, merger, and ringdown. The inspiral phase, characterized by the gradual tightening of the binary system before the merger, contains rich information about the system's parameters and dynamics. 
In our future work, we aim to simultaneously include the inspiral, merger, and ringdown phases in the likelihood function of our detection framework.
Integrating these phases is expected to provide a more comprehensive understanding of the gravitational wave sources, improve parameter estimation accuracy, and enhance the overall sensitivity of the detection method. This comprehensive approach will require the adoption of more sophisticated waveform models, e.g., SEOBNR~\cite{2023PhRvD.108l4035P}, and the adaptation of our optimization algorithms to handle the increased complexity of the combined signal. 

The data that support the findings of this article are openly available~\cite{tao_2025_15024591}.

\acknowledgments
We thank Zhoujian Cao (Beijing Normal University), Siyuan Chen (Shanghai Astronomical Observatory), and Richard Price (MIT) for their helpful discussions.
Y.W. gratefully acknowledges support from the National Key Research and Development Program of China (No. 2023YFC2206702 and No. 2022YFC2205201), the National Natural Science Foundation of China (NSFC) under Grants No. 11973024, Major Science and Technology Program of Xinjiang Uygur Autonomous Region (No. 2022A03013-4), and Guangdong Major Project of Basic and Applied Basic Research (Grant No. 2019B030302001). 
We acknowledge the High Performance Computing Platform at Huazhong University of Science and Technology for providing computational resources. 
The authors thank the anonymous referee for helpful comments and suggestions.

\end{document}